\documentclass[%
reprint,
superscriptaddress,
amsmath,amssymb,
prl
]{revtex4-2}

\usepackage{dcolumn}
\usepackage{bm}

\usepackage{graphicx}
\graphicspath{{./Images/}}

\usepackage{siunitx}

\usepackage{upgreek}

\usepackage{xcolor}

\begin{document}

\title{Imaging signatures of the local density of states in an electronic cavity}

\author{Carolin Gold}
\email[]{cgold@phys.ethz.ch}
\affiliation{Solid State Laboratory, ETH Zurich, 8093 Zurich, Switzerland}

\author{Beat A. Br\"am}
\affiliation{Solid State Laboratory, ETH Zurich, 8093 Zurich, Switzerland}

\author{Michael S. Ferguson}
\affiliation{Institute for Theoretical Physics, ETH Zurich, 8093 Zurich, Switzerland}

\author{Tobias Kr\"ahenmann}
\affiliation{Solid State Laboratory, ETH Zurich, 8093 Zurich, Switzerland}
\author{Andrea Hofmann}
\affiliation{Solid State Laboratory, ETH Zurich, 8093 Zurich, Switzerland}

\author{Richard Steinacher}
\affiliation{Solid State Laboratory, ETH Zurich, 8093 Zurich, Switzerland}

\author{Keith R. Fratus}
\affiliation{Université de Strasbourg, CNRS, Institut de Physique et Chimie des Matériaux de Strasbourg, UMR 7504, F-67000 Strasbourg, France}
\affiliation{HQS Quantum Simulations, Haid-und-Neu-Str.~7, 76131 Karlsruhe, Germany}
\author{Christian Reichl}
\affiliation{Solid State Laboratory, ETH Zurich, 8093 Zurich, Switzerland}

\author{Werner Wegscheider}
\affiliation{Solid State Laboratory, ETH Zurich, 8093 Zurich, Switzerland}

\author{Dietmar Weinmann}
\affiliation{Université de Strasbourg, CNRS, Institut de Physique et Chimie des Matériaux de Strasbourg, UMR 7504, F-67000 Strasbourg, France}

\author{Klaus Ensslin}
\affiliation{Solid State Laboratory, ETH Zurich, 8093 Zurich, Switzerland}
\affiliation{Quantum Center, ETH Zurich, CH-8093 Zurich, Switzerland}
\author{Thomas Ihn}
\affiliation{Solid State Laboratory, ETH Zurich, 8093 Zurich, Switzerland}
\affiliation{Quantum Center, ETH Zurich, CH-8093 Zurich, Switzerland}

\date{\today}

\begin{abstract}
We use Scanning Gate Microscopy to study electron transport through an open, gate-defined resonator in a Ga(Al)As heterostructure.
Raster-scanning the voltage-biased metallic tip above the resonator, we observe distinct conductance modulations as a function of the tip-position and voltage. Quantum mechanical simulations reproduce these conductance modulations and reveal their relation to the partial local density of states in the resonator. Our measurements illustrate the current frontier between possibilities and limitations in imaging the local density of states in buried electron systems using scanning gate microscopy.
\end{abstract}

\maketitle

\section{Introduction}
\label{sec:intro}

Scanning Gate Microscopy (SGM) provides a unique means of investigating local properties of carrier transport in semiconductor nanostructures based on buried two-dimensional electron gases (2DEG)~\cite{ErikssonCryogenicscanningprobe1996,ErikssonEffectchargedscanned1996}.
This imaging technique uses the capacitive coupling between the voltage-biased metallic tip scanned above the sample surface and the electrons in the 2DEG.
Successfully imaged local phenomena and systems in various materials range from disorder-induced or engineered localized states~\cite{BleszynskiScannedProbeImaging2007,FallahiImagingSingleElectronQuantum2005,GildemeisterMeasurementtipinducedpotential2007,WoodsideScannedProbeImaging2002,BachtoldScannedProbeMicroscopy2000,BockrathResonantElectronScattering2001,TansPotentialmodulationscarbon2000,ConnollyTiltedpotentialinduced2011,GarciaScanninggatemicroscopy2013,SchnezImaginglocalizedstates2010}, magnetic focussing of electrons~\cite{AidalaImagingmagneticfocusing2007,BhandariImagingCyclotronOrbits2016}, quantum rings~\cite{PalaLocaldensitystates2008,MartinsImagingElectronWave2007,MartinsScanninggatespectroscopy2013,CabosartRecurrentQuantumScars2017}, quantum Hall edge states \cite{PascherImagingConductanceInteger2014}, ballistic as well as viscous regimes of interacting electron liquids~\cite{BraemScanninggatemicroscopy2018}, to the milestone observation of branched electron flow~\cite{TopinkaImagingCoherentElectron2000,TopinkaCoherentbranchedflow2001,JuraUnexpectedfeaturesbranched2007}.
One major goal of SGM is the experimental measurement of the local density of states in nanostructures. Here, SGM provides a unique opportunity, as only few scanning probe experiments allow to image the local density of states directly, and the most successful technique -- scanning tunneling microscopy -- requires the 2DEG to be accessible at the surface.

The imaging of the local density of states with SGM is well studied in theory~\cite{KolasinskiSimulationsimaginglocal2013a,LyPartiallocaldensity2017a,PalaLocaldensitystates2008} and approached experimentally by a number of pioneering experiments~\cite{MartinsImagingElectronWave2007,SteinacherScanninggateexperiments2018,TopinkaImagingCoherentElectron2000,TopinkaCoherentbranchedflow2001,JuraUnexpectedfeaturesbranched2007,AokiDirectImagingElectron2012,FerryOpenquantumdots2011}.
However, a major obstacle for the experimental realization of scanning gate measurements of truly local electron properties is the invasiveness of the tip-induced potential, which alters the quantum states of interest. 
Imaging local quantum mechanical properties, such as the local density of states, with SGM thus requires weakly invasive tip potentials. 
Unfortunately, the measurement signals obtained in the latter are too weak to be resolved unless strongly confined systems are investigated. These, in turn, lead to a significant loss of spatial resolution of the SGM measurement~\cite{SteinacherScanninggateimaging2016}.

Recently, an accommodation for these two competing requirements of weakly invasive tip-potential and sufficient signal strength was found~\cite{SteinacherScanninggateexperiments2018,RosslerTransportSpectroscopySpinCoherent2015}.
An open resonator structure of intermediate size~\cite{YanInterferenceEffectsTunable2017,YanIncipientsinglettripletstates2018,YanMagnetoresistanceelectroniccavity2018} confines only fundamental one-dimensional cavity modes which can be uniquely identified and addressed~\cite{RosslerTransportSpectroscopySpinCoherent2015}.
This holds true even in slightly less open resonators~\cite{SteinacherScanninggateexperiments2018}.
In the latter, the moderate confinement and size of the cavity allows for SGM-imaging with tip potentials smaller than the Fermi energy and sufficient spatial resolution~\cite{SteinacherScanninggateexperiments2018}. 

In this paper, we demonstrate the correlation between the partial local density of states in such an open resonator structure with the conductance modulations observed in SGM-measurements. 
We scan the voltage-biased metallic SGM tip above the structure for both weakly and strongly invasive tip voltages, and observe distinct conductance modulations in the area of the resonator. 
Quantum mechanical simulations do not only exhibit a good qualitative agreement with the experimental data but also display a correlation between the conductance modulations in the SGM conductance maps and the partial local density of states in the cavity. 
These results show that weakly-invasive SGM provides a tool for measuring direct signatures of the partial local density of states in large two-dimensional electronic-structures.

\section{Sample and Experimental Setup}
\label{sec:sample_setup}
\begin{figure}
	\includegraphics[width=\linewidth]{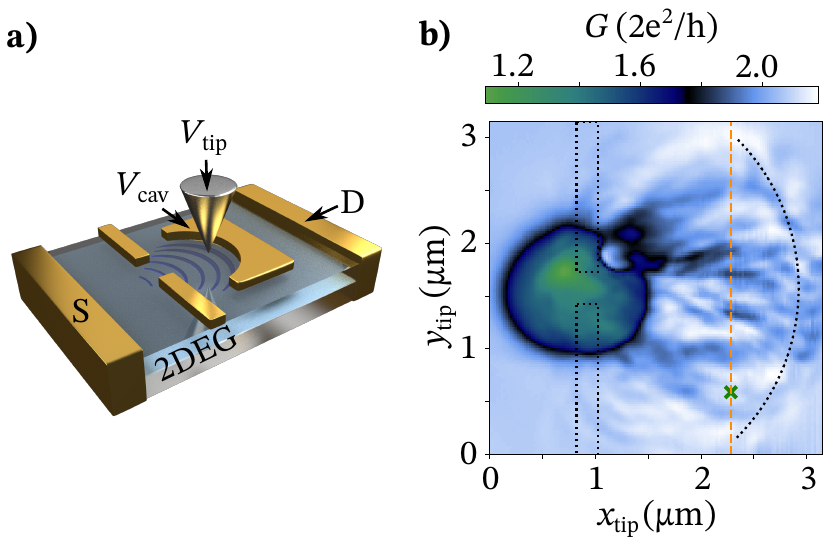}
	\caption{
		(a) Schematic of the measurement setup. Voltage-biased metallic gates (golden) on a Ga(Al)As-heterostructure form an open resonator structure (schematic blue standing wave pattern) in the two-dimensional electron gas (blue). A SGM tip is positioned above the structure to measure the conductance between source (S) and drain (D) ohmic contacts as a function of the tip position.
		(b) SGM image $G(x,y)$ of the cavity area for tip voltage $V_\mathrm{tip}=\SI{-1}{V}$ and cavity gate voltage $V_\mathrm{cav}=\SI{-400}{mV}$. Dotted lines outline the approximate position of the Schottky gates. Measurements in Fig.~\ref{fig2: 3Ddata_linescan_Vcav} are performed along the orange dashed line (Fig.~\ref{fig2: 3Ddata_linescan_Vcav}(b)) respectively the green crossed point (Fig.~\ref{fig2: 3Ddata_linescan_Vcav}(a)). 
	}
	\label{fig1: SEM_and_CavOverview}
\end{figure}


Our measurements are performed at $T=\SI{270}{mK}$ using the measurement setup depicted schematically in Fig.~\ref{fig1: SEM_and_CavOverview}(a). The SGM-tip is raster-scanned above a Ga(Al)As-heterostructure, in which a two-dimensional electron gas (2DEG) resides $\SI{90}{nm}$ underneath the surface. Applying suitable voltages to the lithographically defined metallic Schottky gates allows us to form an open resonator for electrons by depleting the 2DEG underneath the gates.

The back and forth reflection of electrons between the quantum point contact (QPC) through which they are injected into the cavity and the arc-shaped cavity gate enhances the local density of states (LDOS) in the resonator as compared to an open 2DEG in the absence of the cavity. The length ($\SI{2}{\mu m}$) and opening angle ($\SI{90}{^{\circ}}$, defined by the arc of the cavity gate) of the resonator render its area significantly larger than any characteristic lateral size of the SGM-tip induced potential. This avoids the insufficient spatial resolution observed for SGM inside smaller closed structures \cite{SteinacherScanninggateimaging2016}.

Throughout the measurements, the QPC is set to the third conductance plateau (in absence of the cavity) and the resonator is formed by applying a cavity-gate voltage well below the pinch-off voltage. In this configuration, the resonator supports more than 50  populated, radial, spin-degenerate modes~\cite{seeSupplementalMaterial}.

We perform SGM-measurements over the whole cavity area defined by the QPC gates on one side and the cavity gate on the other side. 
By raster scanning the voltage biased tip at a height of $h_{\text{tip}} \approx \SI{80}{nm}$ above the sample surface, we measure the differential conductance $G(x,y)=dI_\mathrm{SD}(x,y)/dV_\mathrm{SD}$ as a function of the tip position $(x,y)$ in a two-terminal setup. 
Here, $I_\mathrm{SD}$ ($V_\mathrm{SD}$) are the current (voltage) between the source and drain ohmic contacts.
The resulting conductance $G(x,y)$ for a weakly-invasive tip (for which the amplitude of the tip-induced potential is much smaller than the Fermi-energy) is depicted in Fig.~\ref{fig1: SEM_and_CavOverview}(b). 
It exhibits conductance modulations in the cavity area, which are in good agreement with previous measurements on the same sample~\cite{SteinacherScanninggateexperiments2018}. 
Cross-capacitance between the QPC and cavity gates, as well as scattering of electrons by the cavity gate back through the QPC reduce the overall conductance well below $G=3 \times 2e^2/h$.

\section{Tip influence on the modulated cavity conductance}

\begin{figure}
	\includegraphics[width=\linewidth]{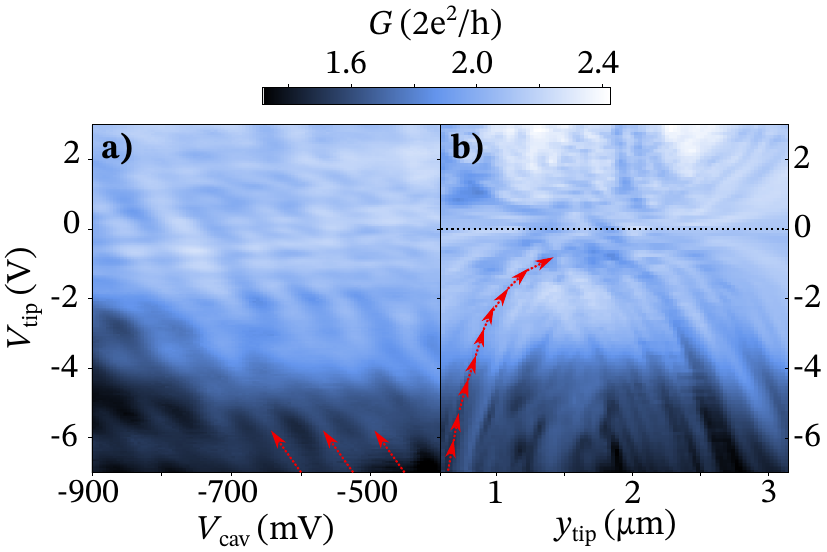}
	\caption{
		\textbf{(a)} Differential conductance $G(V_\mathrm{cav},V_\mathrm{tip})$ as a function of cavity voltage $V_\mathrm{cav}$ and tip voltage $V_\mathrm{tip}$. The tip position is marked by the green cross in Fig.~\ref{fig1: SEM_and_CavOverview}(b).
		\textbf{(b)} Differential conductance $G(y_\mathrm{tip},V_\mathrm{tip})$ along the orange dashed line in Fig.~\ref{fig1: SEM_and_CavOverview}(b).
	}
	\label{fig2: 3Ddata_linescan_Vcav}
\end{figure}

\label{sec: 3D_data}
To study the influence of the tip-induced potential on the cavity, we measure the conductance $G(V_\mathrm{cav},V_\mathrm{tip})$ of the device for various fixed tip positions along the orange dashed line in Fig.~\ref{fig1: SEM_and_CavOverview}(b). 
By tuning the voltage $V_\mathrm{cav}$ in a range well below the pinch-off of the underlying 2DEG, we vary the electronic length of the cavity. 
In contrast, the tip-induced electrostatic potential with maximum value $U_\mathrm{t}$ is varied over the full range from strongly invasive tip potentials ($U_\mathrm{t}> E_\mathrm{F}$, where $E_\mathrm{F}$ is the Fermi energy) to weakly invasive tip potentials ($U_\mathrm{t}<E_\mathrm{F}$)~\cite{SteinacherScanninggateexperiments2018}.
Figure~\ref{fig2: 3Ddata_linescan_Vcav}(a) exemplarily depicts the conductance measured at the tip position marked by the green cross in Fig.~\ref{fig1: SEM_and_CavOverview}(b) ($y_\mathrm{tip}=\SI{0.59}{\mu m}$). 
We observe distinct conductance modulations, which are equidistantly spaced and diagonal as a function of the tip- ($V_\mathrm{tip}$) and cavity- ($V_\mathrm{cav}$) voltages, as indicated by the red arrows.

The Fourier transform of the data in Fig.~\ref{fig2: 3Ddata_linescan_Vcav}(a) reveals a $\lambda_\mathrm{F}/2$-periodicity of the conductance modulations~\cite{seeSupplementalMaterial}. 
This observation is evidence for quasi one-dimensional radial cavity modes~\cite{RosslerTransportSpectroscopySpinCoherent2015}, which 
are separated in energy~\cite{FergusonLongrangespincoherencestronglycoupled2017} by more than $eV_\mathrm{SD}$~\cite{seeSupplementalMaterial}.
Electronic transport through the cavity is modulated by the cavity modes, which are shifted in energy by either the tip-voltage $V_\mathrm{tip}$ or the cavity-gate voltage $V_\mathrm{cav}$. This leads to the diagonal conductance modulations seen in Fig.~\ref{fig2: 3Ddata_linescan_Vcav}(a).

In order to improve our understanding of the relationship between the regular conductance modulations in Fig.~\ref{fig2: 3Ddata_linescan_Vcav}(a) and the apparently random modulations in Fig.~\ref{fig1: SEM_and_CavOverview}(b), we measure the conductance as a function of the tip potential and tip position along the orange dashed line in Fig.~\ref{fig1: SEM_and_CavOverview}(b). 
We plot the resulting conductance $G(y_\mathrm{tip},V_\mathrm{tip})$ in Fig.~\ref{fig2: 3Ddata_linescan_Vcav}(b).
At strongly invasive, negative tip potentials the overall conductance is reduced similar to the observation in Fig.~\ref{fig2: 3Ddata_linescan_Vcav}(a). 
Transitioning from strongly to weakly invasive tip potentials, peaks and troughs in the conductance shift towards the middle of the cavity ($y_\mathrm{tip}=\SI{1.6}{\mu m}$), thus forming arcs. 
Ultimately, the conductance modulations disappear for a small range of tip voltages centered around $V_\mathrm{tip}=\SI{0}{V}$ (c.f. dashed line in Fig.~\ref{fig2: 3Ddata_linescan_Vcav}(b)). 
Increasing the tip voltage beyond this region, we observe a similar but mirrored behavior for conductance modulations at positive tip potentials.

At the common axis in Fig.~\ref{fig2: 3Ddata_linescan_Vcav} ($V_\mathrm{cav}=\SI{-400}{mV}$, $y_{\mathrm{tip}}=\SI{0.59}{\mu m}$), the conductance modulations in  $G(V_\mathrm{cav},V_\mathrm{tip})$ line up with the conductance modulations in $G(y_\mathrm{tip},V_\mathrm{tip})$. 
We conclude that the arced modulations in Fig.~\ref{fig2: 3Ddata_linescan_Vcav}(b) are related to the discrete radial cavity modes.

\section{Understanding the modulated cavity conductance}

There are two common approaches to evaluating the transport properties of non-interacting mesoscopic devices, which have been used extensively to understand SGM measurements: semi-classical expansions~\cite{TopinkaCoherentbranchedflow2001,HellerBranchingFringingMicrostructure2003,PoltlClassicaloriginconductance2016,FratusEnergystabilitybranching2019}, and tight-binding calculations~\cite{PalaLocaldensitystates2008,BoursManipulatingquantumHall2017b,Mrenca-KolasinskaImagingbackscatteringgraphene2017,BrunImagingDiracfermions2019}.
The former is formulated in terms of the classical trajectories of electrons exiting the QPC.
These are then guided onto branches with an increased electron flow due to the random disorder potential generated by the ionized donors~\cite{TopinkaCoherentbranchedflow2001,TopinkaImagingCoherentElectron2000, JuraUnexpectedfeaturesbranched2007}.
The action of a cavity gate leads to a back-folding of the branches, which are simultaneously deflected by the tip-induced potential.
Some of us~\cite{FratusSignaturesfoldedbranches2021} have used a semi-classical approach to theoretically investigate the effect of varying the tip-branch distance and tip voltage, which leads to arc-shaped conductance features in agreement with the experimental observation in Fig.~\ref{fig2: 3Ddata_linescan_Vcav}(b).
Here, we choose instead to use tight-binding calculations, which reproduce both the arc-shaped as well as the $\lambda_\mathrm{F}/2$-periodic conductance modulations in Fig.~\ref{fig2: 3Ddata_linescan_Vcav} and provide an intuitive framework to understand the physical processes in terms of cavity modes.
%

\begin{figure}
	\includegraphics[width=\linewidth]{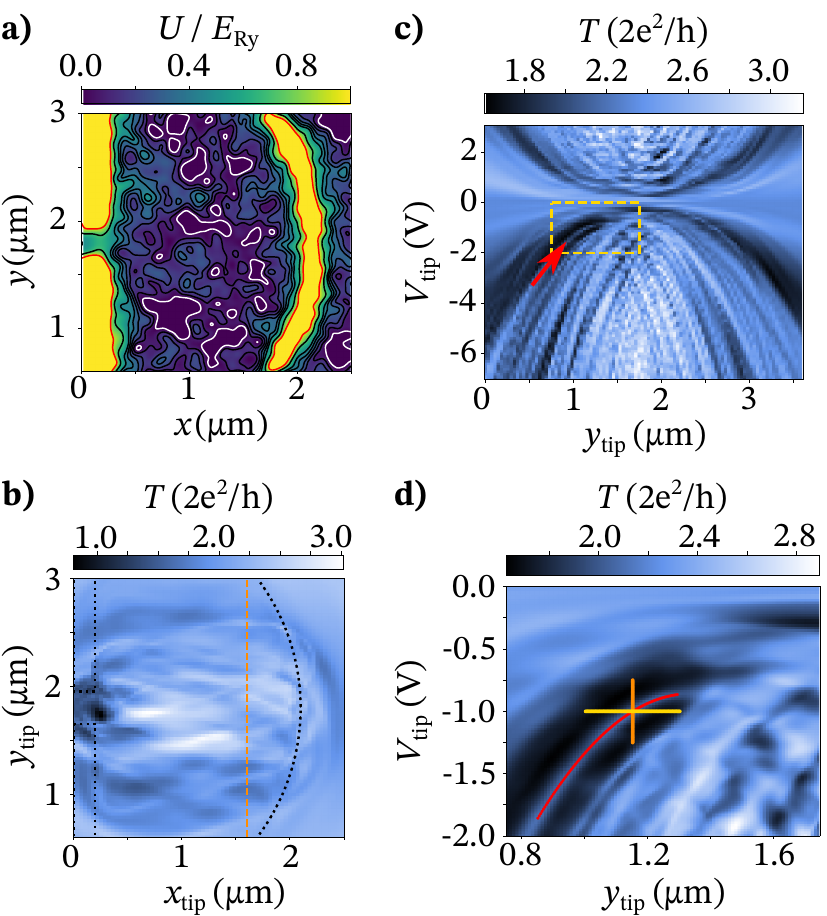}
	\caption{KWANT simulations of the cavity.
		(a) Electrostatic potential $U$ (in units of the Rydberg-Energy) of the cavity in absence of the tip.
		(b) Numerical SGM data of the transmission through the cavity as a function of the tip position ($x_\mathrm{tip},y_\mathrm{tip}$) within the cavity area. 
		(c) Numerical transmission as a function of tip voltage and tip position along the dashed orange line in Fig.~\ref{fig3: Simulation-SGM}(b).
		(d) High-resolution calculation of the transmission within the yellow rectangle in Fig.~\ref{fig3: Simulation-SGM}(c). The fitted maxima of the parabolic feature of interest are denoted by the red parabola. The simulations depicted in Fig.~\ref{fig4: Simulation-LDOS} are performed for the tip parameters denoted by the red arc and the orange/yellow lines.
	}
	\label{fig3: Simulation-SGM}
\end{figure}

Specifically, we perform our tight-binding calculations using the KWANT package~\cite{GrothKwantsoftwarepackage2014}.
The gate geometry in the simulations is similar to the experiment and the potential of the gates is modeled according to the potential suggested in Ref.~\cite{DaviesModelingpatternedtwo1995}. 
The magnitude of the potential applied to the QPC-gates for all simulations is chosen such that the QPC is set to the third conductance plateau in absence of the cavity. 
The disorder in the cavity is modeled by remote impurities positioned at a distance $s=\SI{60}{nm}$ from the two-dimensional electron gas. 
Realizing a specific disorder potential configuration~\cite{seeSupplementalMaterial}, we account for the finite thickness of the electron gas as well as its stand-off distance from the Ga(Al)As-surface by using the Fang-Howard variational wave function. 
Furthermore, we include Thomas-Fermi screening of the disorder potential and assume a spatial correlation of the ionized donors to match the high mobility in our sample~\cite{seeSupplementalMaterial}.
The thus modeled electrostatic potential $U$ (in units of the effective Rydberg-energy $E_\mathrm{Ry}=\SI{5.763}{meV}$) of the system in the absence of the tip is depicted in Fig.~\ref{fig3: Simulation-SGM}(a). 
The tip is included in the simulation via its electrostatic potential induced in the two-dimensional electron gas~\cite{SteinacherScanninggateexperiments2018}. 
Here, we use a (long-ranged) Lorentzian potential motivated by a breadth of previous works~\cite{ErikssonCryogenicscanningprobe1996,PalaLocaldensitystates2008,SteinacherScanninggateinducedeffectsspatial2015,BrunImagingDiracfermions2019} though a Gaussian potential leads to similar results~\cite{seeSupplementalMaterial}.

\begin{figure*}
	\includegraphics[width=\linewidth]{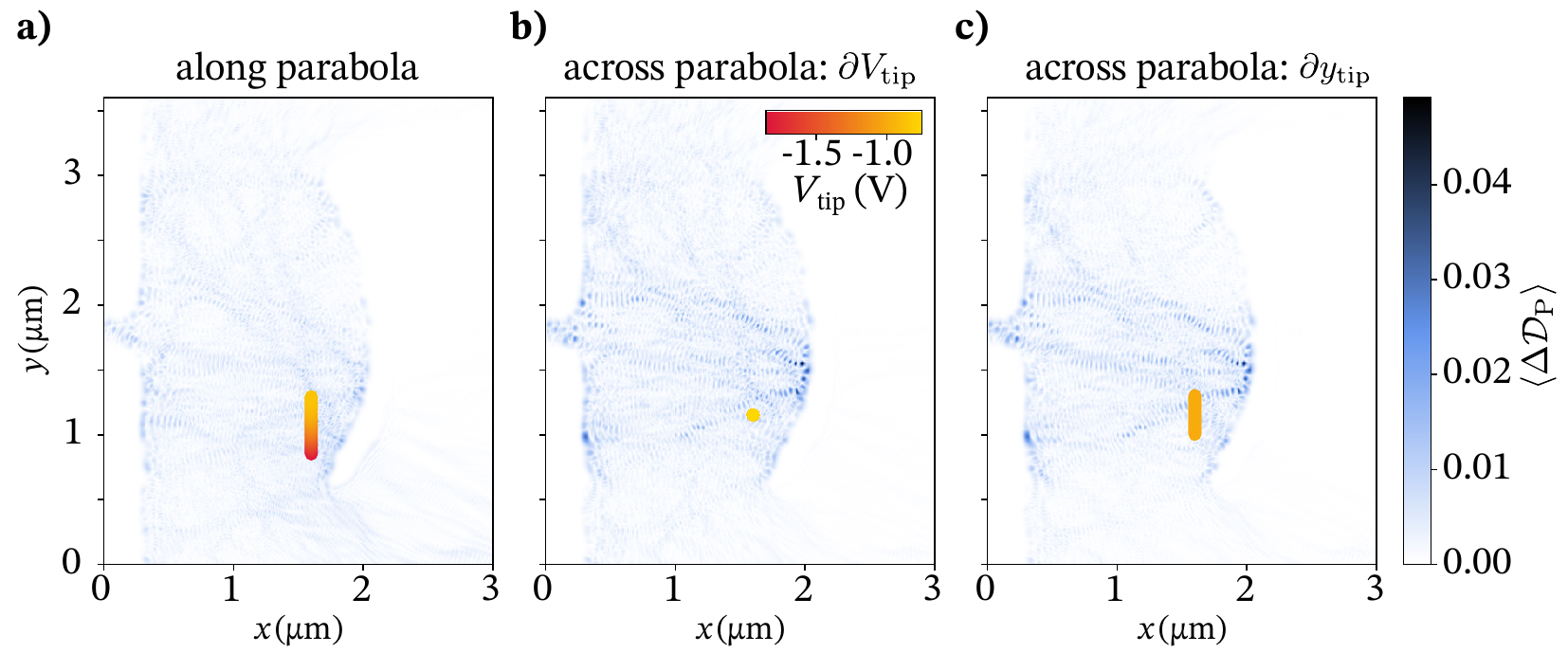}
	\caption{
		Spatially resolved variation of the partial local density of states
		(a) along the conductance maximum,
		(b) across the conductance maximum for fixed tip-position and varying tip-voltage, and
		(c) across the conductance maximum for fixed tip-voltage and varying tip position $y_\mathrm{tip}$.
		The orange-red dots inside the cavity area denote the tip-positions and tip-voltages for the respective data set.
	}
	\label{fig4: Simulation-LDOS}
\end{figure*}


We perform numerical SGM measurements by calculating the transmission between the source and drain contacts at the Fermi energy as a function of the tip position $(x_\mathrm{tip},y_\mathrm{tip})$ and the tip voltage $V_\mathrm{tip}$. 
The calculated equivalents to Figs.~\ref{fig1: SEM_and_CavOverview}(b) and~\ref{fig2: 3Ddata_linescan_Vcav}(b) are shown in Figs.~\ref{fig3: Simulation-SGM}(b) and (c) respectively. 
The numerical and physical experiments show striking similarities, with seemingly random modulations in the two-dimensional scan and more regular arc-shaped modulations in the line cut.


We numerically investigate the correlation between these transmission-modulations and the cavity modes.
To this end, we concentrate on a single conductance maximum, which is isolated from its neighbors and flattens at $y_\mathrm{tip}\approx\SI{1.4}{\mu m}$, see Fig.~\ref{fig3: Simulation-SGM}(d).
We calculate the partial local density of states $\mathcal{D}_\mathrm{P}$ originating from the source contact in the cavity~\footnote{The partial local density of states originating from the source contact is the contribution of the scattering states impinging from that contact to the local density of states, see e.g.~\cite{LyPartiallocaldensity2017,GramespacherLocaldensitiesdistribution1999a,ButtikerBasicElementsElectrical1996}} for each tip-parameter along the conductance maximum and along two cuts of fixed tip potential and position respectively.
The tip-parameters are chosen such that all three lines cross in a single point $(x_\mathrm{tip}^\mathrm{c},y_\mathrm{tip}^\mathrm{c},V_\mathrm{tip}^\mathrm{c})$, see Fig.~\ref{fig3: Simulation-SGM}(d). 
The resulting partial local density of states $\mathcal{D}_\mathrm{P}(x_\mathrm{tip}, y_\mathrm{tip}, V_\mathrm{tip})$ is a two dimensional map in $(x,y)$ which varies as a function of the parameters $x_\mathrm{tip}, y_\mathrm{tip}, V_\mathrm{tip}$~\cite{SeeSupplementalmovies}.

To facilitate the comparison of the three different cuts (red, yellow and orange line in Fig.~\ref{fig3: Simulation-SGM}(d)), we calculate the average (over the length of each cut) deviation in the partial local density of states
\begin{align*}
		\langle\Delta \mathcal{D}_\mathrm{P}\rangle=\langle \left|\mathcal{D}_\mathrm{P}(x_\mathrm{tip}^\mathrm{(i)},y_\mathrm{tip}^\mathrm{(i)},V_\mathrm{tip}^\mathrm{(i)})-\mathcal{D}_\mathrm{P}(x_\mathrm{tip}^\mathrm{c},y_\mathrm{tip}^\mathrm{c},V_\mathrm{tip}^\mathrm{c}) \right| \rangle \\ 
		=\frac{1}{N}\sum_i\left|\mathcal{D}_\mathrm{P}(x_\mathrm{tip}^\mathrm{(i)},y_\mathrm{tip}^\mathrm{(i)},V_\mathrm{tip}^\mathrm{(i)})-\mathcal{D}_\mathrm{P}(x_\mathrm{tip}^\mathrm{c},y_\mathrm{tip}^\mathrm{c},V_\mathrm{tip}^\mathrm{c}) \right|,
\end{align*} 
where the index~$i$ indicates the points along each of the cuts.
We thus obtain in Fig.~\ref{fig4: Simulation-LDOS} three two-dimensional plots of the typical spatially resolved variations of $\mathcal{D}_\mathrm{P}$ along each of the three curves in parameter space.

Along the conductance maximum (Fig.~\ref{fig4: Simulation-LDOS}(a)), $\mathcal{D}_\mathrm{P}$ changes weakly and uniformly across the cavity.
On the other hand, both lines which cross the maximum (Fig.~\ref{fig4: Simulation-LDOS}(b,c)) display strong local changes close to $y_\mathrm{tip}\approx\SI{1.4}{\mu m}$, which indicate a change in the structure of the partial local density of states [cf Supplemental Material \cite{seeSupplementalMaterial}].
In terms of cavity modes, we find that the vertical and horizontal cuts through the conductance maximum result in a change of the mode itself. 
Particularly, we observe that the mode is localized in a region centered around the position of the flattening of the conductance maximum in the simulated SGM measurement in Fig.~\ref{fig3: Simulation-SGM}(d). The size of this region is determined by the influence of the tip on the two-dimensional electron gas.
In contrast to this we probe a single mode by following the conductance maximum.

We conclude that the conductance maxima in our experiment are a function of the tip-induced potential and position and contain information about the \textit{unperturbed} partial local density of states.
This demonstrates that weakly invasive SGM-measurements in sample geometries like this open resonator do preserve properties of the partial local density of states even in the presence of the tip-induced potential.
Furthermore, our data shows that information about the partial local density of states can be resolved also for Lorentzian-shaped tip-induced potentials with a full-width half-maximum of $\SI{250}{nm}$. 

Both the experimental and numerical data show that most modes feature an enhanced local density of states in the central region of the cavity. 
Thus, the tip voltage required to tune a specific mode in this region to the Fermi energy is minimal and the transmission features are almost flat along the arc.

In contrast to this, a small variation in the tip-positions in regions close to the edges of the cavity requires a larger tip voltage difference to tune a certain cavity mode to the Fermi energy.
We thus understand the origin of the arc-shaped modulations observed both in the experimental (Fig.~\ref{fig2:    3Ddata_linescan_Vcav}(b)) as well as the theoretical data (Fig.~\ref{fig3: Simulation-SGM}(c)).

\section{Conclusion}
The scanning gate measurements in the open resonator structure presented in this paper reveal distinct conductance modulations as a function of the cavity-gate voltage $V_\mathrm{cav}$, the tip-position ($x_\mathrm{tip},y_\mathrm{tip}$) and the tip-voltage $V_\mathrm{tip}$. 
Numerical simulations using the KWANT package \cite{GrothKwantsoftwarepackage2014} substantiate the premise that these conductance modulations are related to the cavity modes.
While we cannot measure the local density of states directly, the measurements presented in this paper offer a potential platform to extract information about large ($>\SI{250}{nm}$) scale modulations of the density of states from scanning gate measurements. 
A possible measurement scheme to achieve the latter would be to repeat the measurement presented in Fig.~\ref{fig2: 3Ddata_linescan_Vcav}(b) for all tip-positions $x_\mathrm{tip}$ throughout the cavity. 
Information about the local density of states and the localization of the modes is then contained in the exact trend of the thus measured curves and can potentially be extracted [for details see Supplementary Material~\cite{seeSupplementalMaterial}]. 
This method explores new avenues to gain insights into the partial local density of states in buried electron systems. At the same time, it also illustrates the current frontier between possibilities and limitations in determining the latter via SGM measurements. 

\section{Acknowledgment}
\begin{acknowledgments}
	We thank Peter M\"arki, Thomas B\"ahler as well as the staff of the ETH cleanroom facility FIRST for their technical support. We also acknowledge financial support by the ETH Zurich grant ETH-38 17-2, the Swiss National Science Foundation via NCCR Quantum Science and Technology, the French National Research Agency ANR through projects ANR-11-LABX-0058\_NIE (Labex NIE) and ANR-14-CE36-0007-01 (SGM-Bal) and the University of Strasbourg IdEx program.

	\textit{Author contributions} 
	C.R. and W.W. grew the heterostructure. T.K. and A.H. fabricated the device. R.S. designed the device.
	C.G. performed the measurements with support from B.B. 
	C.G. and M.S.F. developed the interpretation in terms of the partial local density of states. 
	C.G. performed the numerical simulations with support from M.S.F. 
	K.F. and D.W. developed and implemented the semi-classical interpretation. 
	D.W., K.E. and T.I supervised the projects. 
	All authors discussed the results.
\end{acknowledgments}

\bibliography{BIB_CavLDOS_2}
\clearpage
\begin{center}
	\textbf{\large Supplemental Materials: Imaging signatures of the local density of states in an electronic cavity}
\end{center}

\setcounter{equation}{0}
\setcounter{figure}{0}
\setcounter{table}{0}
\setcounter{page}{1}
\makeatletter
\renewcommand{\theequation}{S\arabic{equation}}
\renewcommand{\thefigure}{S\arabic{figure}}

\appendix 
\section{Experimental Details}
\label{sup-sec: expDetails}
\begin{figure}
	\includegraphics[width=\linewidth]{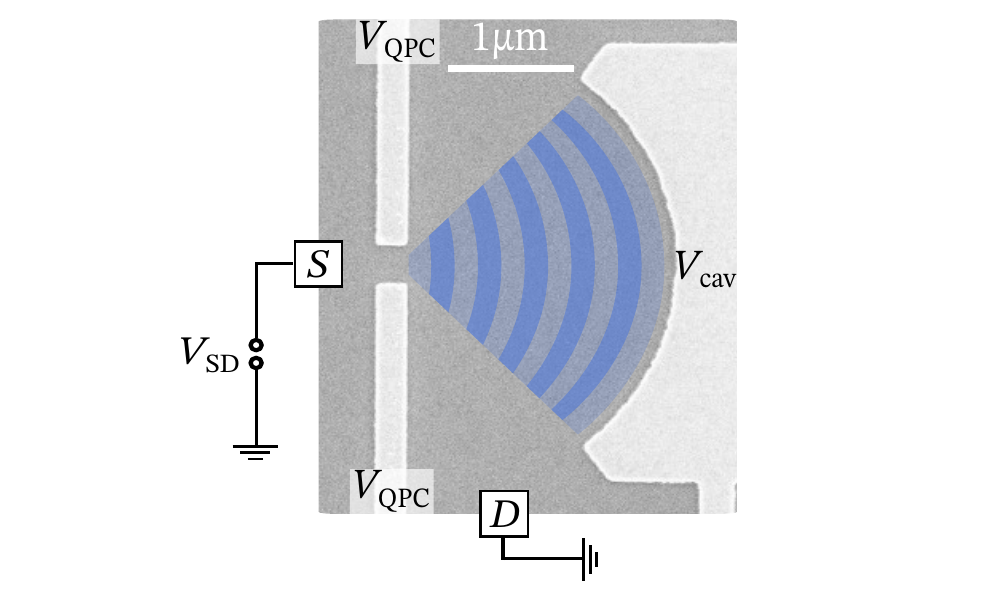}
	\caption{Scanning electron micrograph of the sample. Voltage-biased metallic gates (light gray) on a Ga[Al]As-heterostructure (dark gray) form an open resonator structure (schematic blue standing wave pattern). We measure the conductance through the resonator by voltage biasing the source contact (S) while keeping the drain contact (D) grounded.}
	\label{fig1: SEM}
\end{figure}

A scanning electron micrograph of our open resonator structure is depicted in Fig.~\ref{fig1: SEM}. The two-dimensional electron gas (2DEG) of the GaAs/AlGaAs-heterostructure resides $\SI{90}{nm}$ underneath the surface and has an electron density of $n=\SI{1.9e11}{cm^{-2}}$ and a mobility of $\mu=\SI{4.38e6}{cm^2/Vs}$ at the base temperature of $\SI{270}{mK}$. Lithographically defined metallic Schottky gates (light gray in Fig.~\ref{fig1: SEM}) form an open resonator. The latter consists of an arc-shaped cavity gate with a radius of $\SI{2}{\mu m}$ and an opening angle of $\SI{90}{^{\circ}}$ as well as a $\SI{300}{nm}$ wide quantum point contact (QPC) positioned at the focal point of the arc.

All our measurements are performed by applying an ac-voltage of  $V_\mathrm{SD}=\SI{50}{\mathrm{\mu} V_{rms}}$ to the source contact (S in Fig.~\ref{fig1: SEM}) while keeping the drain contact (D) grounded. The QPC is set to the third conductance plateau ($G=3 \times 2e^2/h
$) in absence of the cavity by applying a voltage of $V_\mathrm{QPC}=\SI{-570}{mV}$ to both QPC gates (Fig.~\ref{fig1: SEM}). In this QPC configuration, the 2DEG underneath the cavity gate can be pinched off by a cavity-gate voltage of $V_\mathrm{cav}=\SI{-240}{mV}$. In order to ensure a fully formed cavity, we choose $V_\mathrm{cav}=\SI{-400}{mV}$, well below the pinch-off voltage. Together with the QPC voltage, this voltage defines the working point for all measurements presented unless stated otherwise.

\section{Cavity-Mode spacing}
\label{sup-sec: cavModes}
\begin{figure}
	\includegraphics[width=\linewidth]{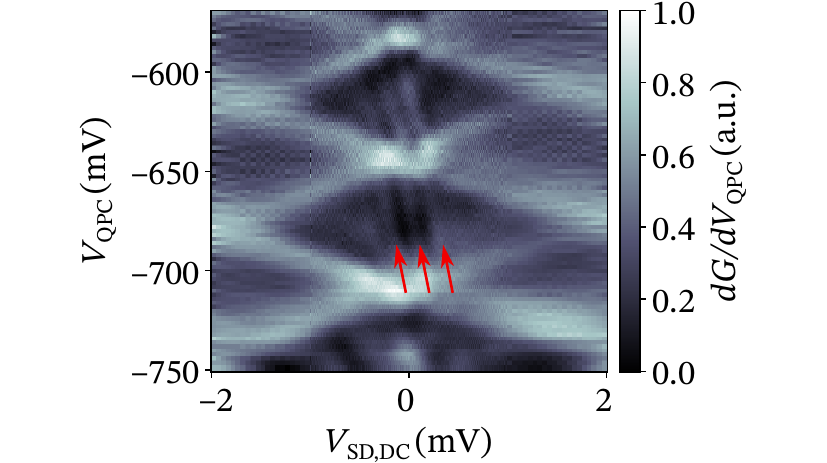}
	\caption{Numerical derivative $dG/dV_\mathrm{QPC}$ of the differential conductance as a function of the dc source-drain voltage $V_\mathrm{SD,DC}$ and the QPC-voltage $V_\mathrm{QPC}$. Red arrows denote the cavity modes. The data for $|V_\mathrm{SD,DC}|>1.0$ is smoothed numerically to improve the visibility of the diamond-shaped patterns despite a lower experimental resolution for these source-drain voltages. A dataset containing this data has also been published in one of our other works~\cite{GoldLocalsignatureselectronelectron2021}.}
	\label{fig1p1: CavModes}
\end{figure}

An important parameter indicating whether individual cavity modes can be resolved is the ratio of the cavity-mode spacing $|\Delta E_\mathrm{modes}|$ and the energy window $\Delta E_\mathrm{SD}$ opened by the source-drain voltage $V_\mathrm{SD}$. The latter can be determined directly as $|\Delta E_\mathrm{SD}|=|e*V_\mathrm{SD}|=\SI{50}{\mu eV}$, while the former has to be determined experimentally using finite-bias spectroscopy. We therefore measure the conductance through the cavity as a function of the dc source-drain voltage $V_\mathrm{SD,DC}$ and the QPC-voltage $V_\mathrm{QPC}$. The numerical derivative of this conductance is depicted in Fig.~\ref{fig1p1: CavModes}. It exhibits the characteristic diamond-shaped pattern associated with non-equilibrium transport through QPCs. Each rhombus of constant differential conductance ($dG/dV_\mathrm{QPC}=\mathrm{const.}$) corresponds to a QPC-plateau. Additionally to this sequence of rhombus patterns, we observe lines of increased $dG/dV_\mathrm{QPC}$ (red arrows), which are only observed for depleting cavity gate voltages. These lines are the experimental signature of nearly equally spaced one-dimensional cavity modes which are shifted through the bias window $\Delta E_\mathrm{SD}$ for varying QPC- and SD-voltages. The bias-spectroscopy allows us to determine a cavity-mode spacing of $|\Delta E_\mathrm{modes}|=\SI[separate-uncertainty = true]{236 \pm 19}{\mu eV}$, which is significantly larger than the bias window $|\Delta E_\mathrm{SD}|$ and the thermal broadening.

\section{Estimating the number of populated cavity modes and their energy spacing}
\label{sup-sec: Number_CavityModes}

The number of populated modes supported by a one-dimensional cavity can be estimated as $N_\mathrm{cav}=2L_\mathrm{cav}/\lambda_\mathrm{F}$, where $L_\mathrm{cav}$ is the length of the cavity and $\lambda_\mathrm{F}$ is the Fermi wavelength. To estimate the length $L_\mathrm{cav}$ of the cavity at certain gate voltages, we consider the lithographic size of the cavity ($L_\mathrm{l}$) and the change $\Delta L_\mathrm{cav}$ of the depletion length of the gates as a function of the applied voltages. For the cavity gate, the latter can be approximated by a linear relation with slope $P_\mathrm{cav}=\SI{350}{nm/V}$ \cite{SteinacherScanninggateinducedeffectsspatial2015}. A similar ratio can be estimated for the QPC gates by considering the voltage difference between the onset of 2DEG depletion and the complete pinch-off of the QPC. In our sample, the latter is given as $\Delta V_\mathrm{QPC}=|\SI{750}{mV}-\SI{290}{mV}|=\SI{460}{mV}$ for each QPC-gate. As each of the gates only needs to pinch-off half of the QPC width of $L_\mathrm{QPC}=\SI{300}{nm}$, the depletion length of the QPC gates changes with $P_\mathrm{QPC}=\frac{\SI{150}{nm}}{\SI{460}{mV}}=\SI{326}{nm/V}$. With this, we can estimate the length of the cavity at certain gate voltages as $L_\mathrm{cav}=L_\mathrm{l}-\Delta L_\mathrm{cav}$. We estimate $\Delta L_\mathrm{cav}$ via the change of the depletion length of each gate as a function of the gate voltage using the proportionality constants ($P_\mathrm{cav},P_\mathrm{QPC}$) and the voltage-difference between the applied gate voltage and the pinch-off voltage of the respective gate. The estimated cavity length at the working point is thus $L_\mathrm{cav}=\SI{1.853}{\mu m}$. Together with a Fermi-wavelength of $\lambda_\mathrm{F}=\sqrt{2\pi/n_\mathrm{s}}$, this yields $N_\mathrm{cav}=2L_\mathrm{cav}/\lambda_\mathrm{F}=\SI{64}{modes}$.

For resonators supporting such large numbers of one-dimensional, populated modes, the energy spectrum of the radial cavity modes is similar to the spectrum of a one-dimensional box with length L,
\begin{equation*}
	E_\mathrm{n}=\frac{\pi^2\hbar^2n^2}{2m^{\ast}L^2}.
\end{equation*}
Here, n is the quantum number of the mode, $m^{\ast}$ the effective mass of the particles in the box and $L$ is the radius of the cavity for the radial modes confined in our resonator.
The energy spacing of the modes at the Fermi-energy $E_\mathrm{F}$ is then given as
\begin{equation*}
	\Delta E_\mathrm{modes}=\frac{\pi\hbar\sqrt{2E_\mathrm{F}}}{\sqrt{m}L}.
\end{equation*}
This allows us to estimate a cavity mode energy spacing of $\Delta E_\mathrm{modes}=\SI{209}{\mu eV}$, which is in good agreement with the experimentally measured value of $\SI[separate-uncertainty = true]{236\pm 19}{\mu eV}$.

\section{Fourier Transform of the conductance modulations\\ in $V_\mathrm{tip}$ and $V_\mathrm{cav}$.}
\label{sup-sec: FFT_condModul}
\begin{figure}
	\includegraphics[width=\linewidth]{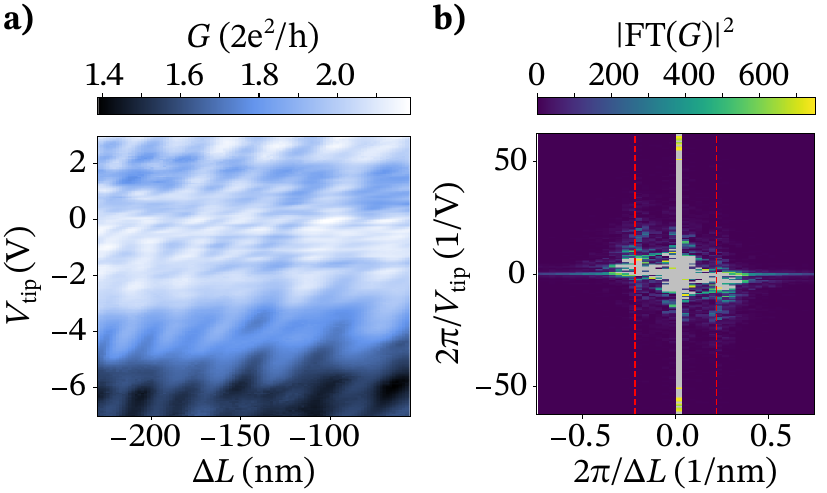}
	\caption{Periodicity of the conductance modulations. (a)~Conductance $G$ as a function of tip-voltage $V_\mathrm{tip}$ and cavity-gate induced change in the cavity length $\Delta L$. (b) Fourier transformation of the conductance in Fig.~\ref{fig2: FFT}(a). The red dashed lines denote twice the Fermi-wavenumber $\pm 2k_\mathrm{F}= \pm 2\pi/(\lambda_\mathrm{F}/2)$. The values depicted in grey exceed the upper limit of the color scale. }
	\label{fig2: FFT}
\end{figure}

In order to study the periodicity of the conductance modulations as a function of the tip-voltage and cavity-gate voltage depicted in Fig.~2(a), we calculate the Fourier transform of this conductance. The difference $\Delta V_\mathrm{cav}$ between the cavity gate voltage and the pinch-off voltage of the latter changes the cavity length by $\Delta L = P_\mathrm{cav}V_\mathrm{cav}=\SI{350}{nm/V}\times V_\mathrm{cav}$. This allows us to transform the data depicted in Fig.~2(a) into the conductance $G(\Delta L, V_\mathrm{tip})$ depicted in Fig.~\ref{fig2: FFT}(a). Calculating the Fourier transform of the latter yields the data depicted in Fig.~\ref{fig2: FFT}(b). Here, $|\mathrm{FT}(G)|^2$ is the power spectral density of the Fourier transform of the conductance $G(\Delta L, V_\mathrm{tip})$ (cf Fig.~\ref{fig2: FFT}(a)) which is depicted as a function of $2\pi/\Delta L$ and $2\pi/V_\mathrm{tip}$. The red dashed lines denote twice the Fermi wavevector $\pm 2k_\mathrm{F}= \pm 2\pi/(\lambda_\mathrm{F}/2)\approx\pm \SI{0.22}{1/nm}$ and coincide with the maxima of the Fourier transform $|\mathrm{FT}(G)|^2$. The conductance modulations in Fig.~2(a) and Fig.~\ref{fig2: FFT}(a) thus have a periodicity of $\lambda_\mathrm{F}/2$.

\section{Implementation of the smooth disorder potential}
\label{sup-sec: Disorder}
Transport properties of Ga(Al)As heterostructures are strongly influenced by various scattering mechanisms present in the sample. Particularly, it has been shown that elastic scattering off remote ionized dopants not only influences the electron mobility of the two-dimensional electron gas at low temperatures~\cite{IhnSemiconductorNanostructuresQuantum2009}, but also introduces small-angle scattering events due to weak, random interactions~\cite{TopinkaCoherentbranchedflow2001,HellerBranchingFringingMicrostructure2003}. We include elastic impurity scattering into our simulations by assuming that the disorder is caused by randomly distributed, singly-ionized dopants in a doping plane at a distance $s$ from the two-dimensional electron gas. Considering Thomas-Fermi screening, the finite thickness of the two-dimensional electron gas via the Fang-Howard variational approach, and spatial correlations of the donor distribution, the screened electrostatic potential in the electron gas depends on a number of parameters. These are the relative dielectric constant $\epsilon$ of Ga(Al)As, the spacer thickness $s$, the two-dimensional Thomas-Fermi wave vector $q_\mathrm{TF}$, the Fang-Howard variational parameter $b$, and for the correlation term, the reciprocal screening length $q_0=2\pi N_\mathrm{d}e^2/(\epsilon \epsilon_0 k_\mathrm{B} T_0)$. In the latter, $N_\mathrm{d}$ is the average area concentration of the charged donors, and $T_0$ the freeze-out temperature. With this, the spatially varying electrostatic potential induced in the two-dimensional electron gas due to the ionized donors can be written as
\begin{multline}
	V\left(\vec{r}\right)=-\frac{e^2}{2\epsilon\epsilon_0}\int \frac{dq^2}{(2\pi)^2}\frac{e^{-qs}}{q+q_\mathrm{TF}} \left(\frac{b}{b+q}\right)^3  \\ \times C_\mathrm{uc}\left(\vec{q}\right)\sqrt{\frac{q}{q+q_0\left(1-e^{-2qs}\right)}}e^{-i\vec{q}\vec{r}}.
	\label{eq:DisorderPot}
\end{multline}
Here, we have further introduced the two-dimensional wave vector $\vec{q}$ with absolute value $q$ and the Fourier transform $C_\mathrm{uc}\left(\vec{q}\right)$ of the fluctuating part of the uncorrelated random distribution of donors.

In the following, we will outline the key steps leading to this disorder potential.

\paragraph{External fluctuation potential:}
First, we assume randomly placed dopants in the doping plane with mean doping density $N_\mathrm{d}$. Considering only the fluctuating part $C\left(\vec{r}\right)$ of the dopant distribution, we can write the (external) fluctuation potential induced in the 2DEG at position $z\geq 0$  by the randomly placed donors in real space as
\begin{equation*}
	V_\mathrm{ext}(\vec{r},z)=-\frac{e^2}{4\pi\epsilon\epsilon_0}\int d\vec{r'}\frac{C(\vec{r'})}{\sqrt{(\vec{r}-\vec{r'})^2+(z+s)^2}},
\end{equation*}
where $\vec{r},\vec{r'}$ are two-dimensional vectors parallel to the plane of the electron gas (see Ref.~\onlinecite{IhnSemiconductorNanostructuresQuantum2009}, Chapter 9.5.1). Here, the plane $z=0$ is chosen as the location of the heterointerface between the AlGaAs barrier and the GaAs.
The Fourier-transform of this potential yields 
\begin{equation*}
	V_\mathrm{ext}(\vec{q},z)=-\frac{e^2}{2\epsilon\epsilon_0}C(\vec{q})\frac{e^{-q|z+s|}}{q},
\end{equation*}
for $q>0$.

\paragraph{Thomas-Fermi-screening:}
Second, we include screening of this external potential by the two-dimensional electron gas. Considering long-range potentials ($2k_\mathrm{F}s\gg1$, where $k_\mathrm{F}$ is the Fermi-wave number) in the quantum limit, we use the Thomas-Fermi approximation and obtain the relation 
\begin{equation*}
	\left<V_\mathrm{tot}(\vec{q})\right>=\left<V_\mathrm{ext}(\vec{q})\right>\frac{q}{q+q_\mathrm{TF}},
\end{equation*}
where $q_\mathrm{TF}$ is the two-dimensional Thomas-Fermi vector (see Ref.~\onlinecite{IhnSemiconductorNanostructuresQuantum2009}, Chapter 9.5.3). Here, the notation $\langle\ldots\rangle$ denotes the expectation value of the respective quantity evaluated with the ground-state wave function $\varphi_0(z)$ in $z$-direction of the two-dimensional electron gas.

\paragraph{Fang-Howard variational approach}
We account for the finite thickness of the electron gas by using the Fang-Howard variational wave function for $\varphi_0(z)$ (see Ref.~\onlinecite{IhnSemiconductorNanostructuresQuantum2009}, Chapter 9.4), thus obtaining
\begin{equation}
\left<V_\mathrm{tot}(\vec{q})\right>=-\frac{e^2}{2\epsilon\epsilon_0}C(\vec{q})\frac{e^{-2qs}}{q+q_\mathrm{TF}}\left(\frac{b}{b+q}\right)^3.
\label{eq: disorderPot_noCorr}
\end{equation}
Here, the Fang-Howard variational parameter $b$, minimizing the total energy of the electron gas in the absence of disorder, is given as a function of the electron density $n_\mathrm{s}$ in the 2DEG and the effective Bohr radius $a_\mathrm{B}^\ast$ according to $b=\left[33\pi n_\mathrm{s}/(2a_\mathrm{B}^\ast)\right]^{1/3}$.

\paragraph{Correlated donor distribution:}
Scattering of electrons off the ionized dopants can be described in first order perturbation theory by Fermi's golden rule according to
\begin{equation*}
	\frac{1}{\tau}=\frac{2\pi}{\hbar}\frac{1}{A^2}\sum_q \left<|V(\vec{q})|^2\right>\delta \left[E(\vec{k})-E(\vec{k}+\vec{q})\right]
\end{equation*} (see Ref.~\onlinecite{IhnSemiconductorNanostructuresQuantum2009}, Chapter 10.7.). However, it has been shown both theoretically~\cite{LassnigRemoteionscattering1988,EfrosMaximumlowtemperaturemobility1990,MonroeIntersiteCoulombrepulsion1991} as well as experimentally~\cite{ColeridgeSmallanglescatteringtwodimensional1991,WilamowskiAppearancedestructionspatial1991,MaudeEffectspatialcorrelation1992,CozSpatialcorrelationsDX1993,BuksCorrelatedchargeddonors1994} that spatial correlations between the ionized donors play an important role in finding mobility values $\mu=|e|\tau/m^\ast$ matching the high mobilities measured experimentally. While there exist different approaches~\cite{KawamuraSpatialcorrelationeffect1996, LassnigRemoteionscattering1988, EfrosDensitystates2D1988, MycielskiFormationsuperlatticeionized1986} to consider the correlation of the donors, we use the approach by Efros et al.~\cite{EfrosDensitystates2D1988,EfrosMaximumlowtemperaturemobility1990}, who introduced the donor correlation according to
\begin{equation*}
	\left<\Tilde{C}(\vec{q})\Tilde{C}(\vec{q'})\right>_\mathrm{average}=N_\mathrm{d}\delta(\vec{q}+\vec{q'})\frac{q}{q+q_0\left(1-e^{-2qs}\right)}.
\end{equation*} 
We thus obtain $\Tilde{C}\left(\vec{q}\right)=C_\mathrm{uc}\left(\vec{q}\right)\sqrt{q/\left(q+q_0\left(1-e^{-2qs}\right)\right)}$ for the Fourier transform of the fluctuating part of the correlated donor distribution. Inserting $\Tilde{C}(\vec{q})$ into Eq.~\eqref{eq: disorderPot_noCorr} for $C(\vec{q})$ and calculating the inverse Fourier transform of the resulting expression then yields the description of the donor-induced electrostatic potential in the 2DEG given in Eq.~\eqref{eq:DisorderPot}.

\subsection{Implementation of the random disorder potential on a discrete lattice:}
\paragraph{Statistical properties of $C_\mathrm{uc}(\vec{q})$.}
We assume a sample of dimensions $L\times W$ in which a large number $M=N_\mathrm{d}LW$ of dopants is distributed uniformly in the whole area of the doping plane. Based on the statistical properties of $C_\mathrm{uc}(\vec{q})$ and owing to the central limit theorem, it can then be shown that the distribution $C_\mathrm{uc}(\vec{q})$ can be written as 
\begin{align}
	C_\mathrm{uc}(\vec{q})&=\sqrt{\frac{N_\mathrm{d}LW}{2}}\left[\alpha(\vec{q})+i\beta(\vec{q})\right]\nonumber \\&=\sqrt{\frac{N_\mathrm{d}LW}{2}}\rho(\vec{q})e^{i\phi(\vec{q})}
	\label{eq:corr}
\end{align}
at any value of $q$.
Here, the real (imaginary) part $\alpha$ ($\beta$) is distributed normally with mean zero and variance $\sigma^2=N_\mathrm{d}LW/2$. Furthermore, $\rho=\sqrt{\alpha^2+\beta^2}$ is the absolute value and $\phi=\angle(\alpha+i\beta)$ the angle of the random complex number $\alpha+i\beta$.

\paragraph{Discretization of the potential on a lattice:}
We then discretize the plane of the electron gas into a grid with lattice constant $a$. The resulting discretization in $\vec{q}$-space has step widths $\Delta q_x=2\pi/L$ and $\Delta q_y=2\pi/W$ and a maximum $q$-value $q_\mathrm{max}=2\pi/a$. Labeling the discrete values $\vec{q}_j$ such that $\vec{q}_0=0$ and  $\vec{q}_{-j}=-\vec{q}_j$ and using Eq.~\eqref{eq:corr},  we can thus discretize the potential $V(\vec{r})$ in Eq.~\eqref{eq:DisorderPot} in real space according to 
\begin{multline*}
	V\left(\vec{r}_k\right)=-\frac{e^2}{2\epsilon\epsilon_0} \frac{\Delta q_x\Delta q_y}{(2\pi)^2}\sum_{j> 0} \frac{e^{-q_js}}{q_j+q_\mathrm{TF}} \left(\frac{b}{b+q_j}\right)^3  \\ \times \sqrt{\frac{q_j}{q_j+q_0\left(1-e^{-2q_js}\right)}}\rho_j\cos\left(\vec{q}_j\vec{r}_k-\phi_j\right)
\end{multline*}
To implement this potential in our simulations, we draw random values of real ($\alpha$) and imaginary ($\beta$) parts from a normal distribution for each wave vector $\vec{q}_j$.

\section{Numerical SGM measurements for a Gaussian-shaped tip potential}
\label{sup-sec: GaussianTip}
\begin{figure}
	\includegraphics[width=\linewidth]{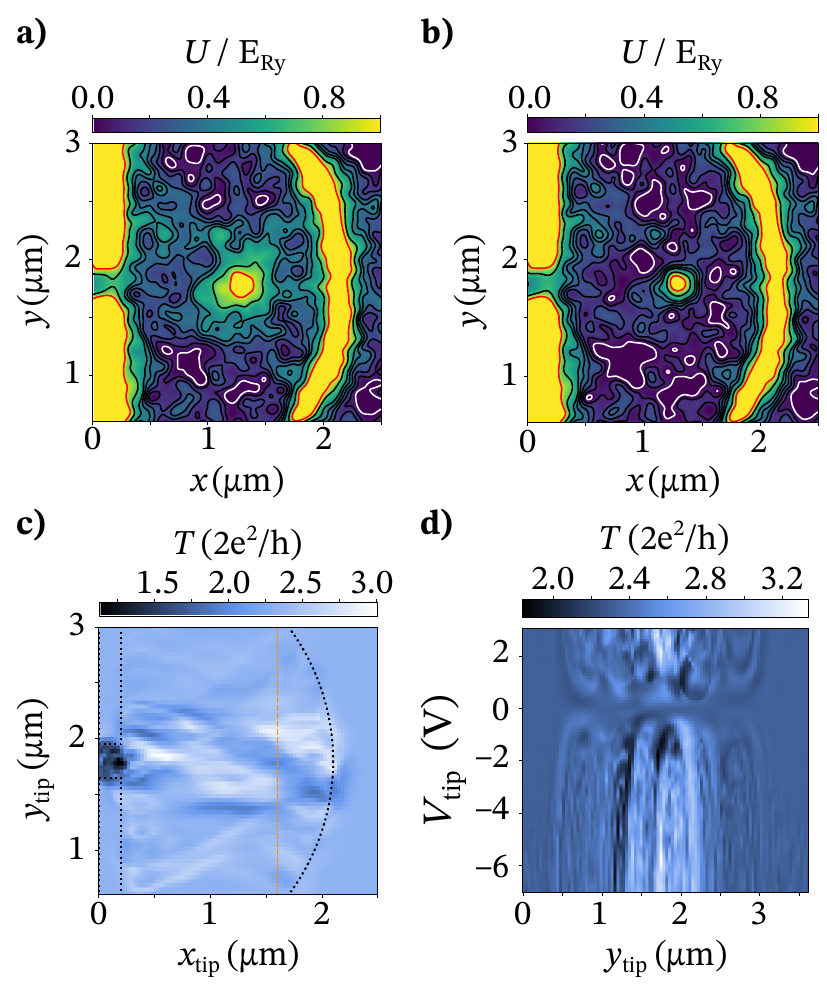}
	\caption{Influence of the shape of the tip-induced potential.
		(a) Potential landscape of the cavity for a (long-ranged) Lorentzian tip potential.
		(b) Potential landscape of the cavity in the presence of a (short-ranged) Gaussian-shaped tip potential.
		(c) Numerical SGM measurement of the transmission through the cavity for a Gaussian tip potential.
		(d) Numerical transmission as a function of the tip-voltage and tip-position along the orange-dashed line in Fig.~\ref{fig3: GaussianTip}(c).
	}
	\label{fig3: GaussianTip}
\end{figure}

The exact shape of the tip potential has a profound influence on the potential landscape within the cavity. 
Using the same numerical approach described in the main text, we compare the influence of these two tip-potential shapes in Fig.~\ref{fig3: GaussianTip}(a) and (b). Here, the tip is positioned at exactly the same position within the cavity and features the same full width at half-maximum for both the Lorentzian and the Gaussian tip potential. The tip-voltage is chosen close to the depletion voltage. The long-range tails of the Lorentzian tip potential influence a large area of the potential landscape within the cavity, while the Gaussian-shaped tip potential induces a more local change. 
The calculated equivalents for the numerical SGM measurements in Fig.~3(b),(c) for a Gaussian-shaped tip induced potential are depicted in Fig.~\ref{fig3: GaussianTip}(c), (d). 
Although we observe arc-shaped conductance modulations in $G(y_\mathrm{tip},V_\mathrm{tip})$ for both the Lorentzian (Fig.~3(c)) and Gaussian (Fig.~\ref{fig3: GaussianTip}(d)) shaped tip potential, we also note several differences between the two cases. 
The Lorentzian-shaped tip-potential results in rather smooth arcs, while the Gaussian tip potential induces narrower arcs with a significantly steeper slope in the outer regions and almost flat maxima at the vertices of the arcs. 
These differences can be understood in terms of the rather long-range tails of the Lorentzian tip in contrast to the short-ranged Gaussian potentials. This consequence is also explained by a classical model calculation based on the deflection of electron trajectories by the tip potential~\cite{FratusSignaturesfoldedbranches2021}.

\section{Statistics of the average difference $\langle \Delta D_\mathrm{P}\rangle$ of the partial local density of states}
\label{sup-sec: statistics}
\begin{figure}

	\includegraphics[width=\linewidth]{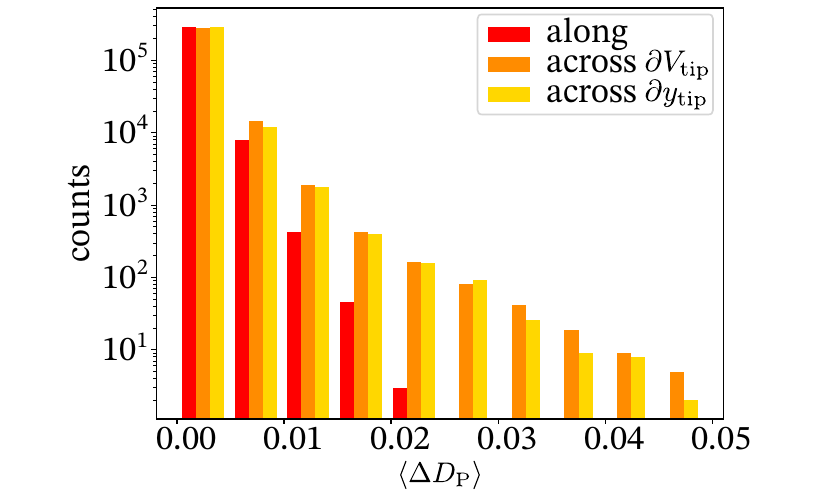}
	\caption{Histogramm of the average difference of the partial local density $\langle \Delta D_\mathrm{P}\rangle$ depicted in Fig.~4 (main text) along (red) and across (orange, yellow) the conductance maximum in Fig.~3(d) (main text). The colors are chosen to match the respective cuts in Fig.~3(c).
	}
	\label{Sfig: PLDOS_Hist}
\end{figure}

To evaluate the data depicted in Fig.~4 of the main text quantitatively, we depict the approximate distribution of the data in form of a histogram in Fig.~\ref{Sfig: PLDOS_Hist}. Here, the red bars correspond to the data along the conductance maximum [cf Fig.~4(a)], whereas the remaining bars correspond to the data across the conductance maximum for varying tip-voltage $\partial V_\mathrm{tip}$ [orange, corresponding to the data in Fig.~4(b)] respectively tip-position $\partial y_\mathrm{tip}$ [yellow, cf data in Fig.~4(c)]. We observe that the distribution of the data points in Fig.~4(a) along the conductance maximum [red line in Fig.~3(d)] is concentrated on average partial local densities $\langle \Delta \mathcal{D}_\mathrm{P}\rangle <0.025$. On the contrary, both data sets obtained across the conductance maximum [cf orange and yellow line in Fig.~3(d)] have significantly higher occurrences of average partial local densities $\langle\Delta \mathcal{D}_\mathrm{P}\rangle$ of up to 0.05. This is in good agreement with our observation that $\mathcal{D}_\mathrm{P}$ changes weakly along the conductance maximum, whereas it changes significantly for both varying tip-position and -voltage across the conductance maximum.

\section{Possible measurement to extract the local density of states from SGM-measurements}
\label{sup-sec: SMG_tomography}
\begin{figure}
	
	\includegraphics[width=\linewidth]{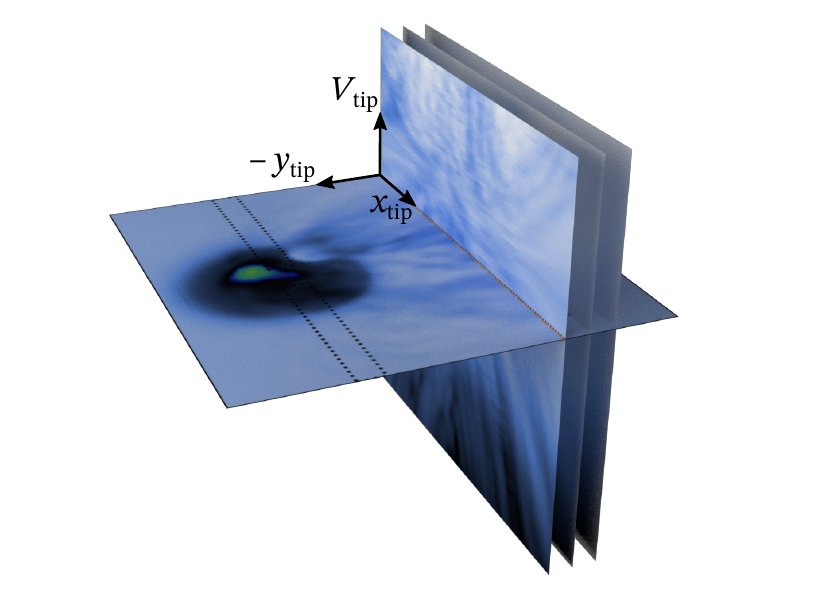}
	\caption{Schematic of a possible measurement to reconstruct the partial local density of states from SGM-measurements. The colormap corresponds to the conductance through the cavity as a function of the tip-position and --voltage.
	}
	\label{Sfig: Tomography}
\end{figure}

While the detailed description and implementation of an experiment allowing one to extract the local density of states from SGM-measurements is beyond the scope of this work, we will give an overview of a possible, yet very elaborate experiment in the following. While such a measurement has been a strong motivator for SGM-measurements since its discovery in 1996~\cite{ErikssonCryogenicscanningprobe1996,ErikssonEffectchargedscanned1996,TopinkaImagingCoherentElectron2000,TopinkaCoherentbranchedflow2001,MartinsImagingElectronWave2007,JuraUnexpectedfeaturesbranched2007,FerryOpenquantumdots2011,AokiDirectImagingElectron2012,SteinacherScanninggateexperiments2018}, our work is a significant milestone to achieving this goal.
Repeating the measurement of the conductance $G(y_\mathrm{tip},V_\mathrm{tip})$ for all tip-positions $x_\mathrm{tip}$ in the cavity yields the conductance $G(x_\mathrm{tip},y_\mathrm{tip},V_\mathrm{tip})$. A schematic overview of such a data-set is depicted in Fig.~\ref{Sfig: Tomography}. The evolution of the conductance maxima as a function of all tip-parameters $(x_\mathrm{tip},y_\mathrm{tip},V_\mathrm{tip})$ contains information about the partial local density of states $\mathcal{D}_\mathrm{P}$. Given such a complete data set and assuming weakly invasive potentials, it should be possible to reconstruct $\mathcal{D}_\mathrm{P}$ up to the resolution given by the width of the tip potential. 

Our proof of principle experiment captures information about $\mathcal{D}_\mathrm{P}$ by measuring the conductance $G(x_\mathrm{tip},V_\mathrm{tip})$ along a line across the cavity. 
The observed conductance maxima correspond to specific cavity modes at the Fermi-energy, which are localized in the cavity and have a finite lifetime.
If higher-order perturbations are negligible, the energy of a distinct cavity mode can thereby be shifted with respect to the Fermi-energy $E_\mathrm{F}$ according to 
\begin{equation}
\Delta E=\int dx\int dy \mathcal{D}_\mathrm{P}(x,y)(V_\mathrm{tip}(x,y)-V_\mathrm{tip}^\mathrm{li}),
\label{eq:Delta_E_cavitymode}
\end{equation}
in first-order perturbation theory. Here, $V_\mathrm{tip}^\mathrm{li}$ is the least-invasive tip voltage. Our numerical observation that the partial local density of states remains relatively invariant along a conductance maximum associated with a cavity mode indicates that such an approach is justified.
Assuming a delta-shaped tip-induced potential for a proof of principle, we obtain a first approximation of this behaviour encoded by
\begin{align*}
	(V_\mathrm{tip}-V_\mathrm{least-invasive}) \cdot \mathcal{D}_\mathrm{P}^\mathrm{i}(x)  \approx const.
\end{align*}
along the maximum labeled by i.
Solving for the linear density of states we obtain $\mathcal{D}_\mathrm{P} = const./V_\mathrm{tip}$. This indicates a maximal density of states of the cavity mode when the trajectory of a maximum is close to the minimally invasive potential, and a reduced $\mathcal{D}_\mathrm{P}$ when the trajectory enters the strongly invasive regime. In both our experiment and theory, we thus observe a maximal density of states for the cavity modes in the cavity and a reduced density of states further away.
The realistic description of our experiment however includes an approximately Lorentzian-shaped tip-induced potential as well as a local density of states that is extended over the entire sample. 
Equation~\eqref{eq:Delta_E_cavitymode} thus becomes a convolution of the spatially extended tip-induced potential and local density of states. Solving for the local density of states thus becomes a deconvolution instead of an inversion but remains in principle feasible.

To simplify the mathematical procedure, we propose a slightly modified sequence of three experimental steps to measure the density of states of a single cavity mode.
First, the tip is centred above the cavity and a weakly-invasive tip potential is applied. Second, the isolated (in energy) cavity level of interest is tuned to the Fermi-energy by changing the cavity gate voltage. Finally, the tip is scanned above the cavity and the resulting shift of the isolated cavity level of interest with respect to the Fermi-energy is compensated for at each tip-position $(x_\mathrm{tip},y_\mathrm{tip})$ by tuning the Fermi-energy of the sample, e.g. via a voltage-biased backgate. The resulting surface of maximal conductance as a function of both $(x_\mathrm{tip},y_\mathrm{tip})$ and $\Delta E_\mathrm{F}$ is directly proportional to the density of states of the cavity mode convoluted with the tip potential. As the resolution of the reconstruction of $\mathcal{D}_\mathrm{P}$ is limited by the width of the tip-induced potential, achieving sharper tip potentials (with widths below the Fermi-wavelength) remains an important and interesting technological challenge, which has not yet been achieved.

\end{document}